# THE EQUIVALENCE PRINCIPLE

# AS A PROBE FOR HIGHER DIMENSIONS

**Paul S. Wesson**

Departments of Physics and Applied Mathematics

University of Waterloo, Waterloo, Ontario N2L 3G1, Canada

Summary: Higher-dimensional theories of the kind which may unify gravitation with particle physics can lead to significant modifications of general relativity. In five dimensions, the vacuum becomes non-standard, and the Weak Equivalence Principle becomes a geometrical symmetry which can be broken, perhaps at a level detectable by new tests in space.

Correspondence: mail to address above, fax= (519) 746-8115, phone= (519) 885-1211 ext. 2215

The Equivalence Principle as a Probe for Higher Dimensions

Objects accelerate at the same rate in a gravitational field even though they have different masses and other physical properties. This is a consequence of the Weak Equivalence Principle, which underlies general relativity.

The wish to unify gravity with the interactions of particle physics has led to a plethora of theories in which Einstein's spacetime is extended to more than 4 dimensions. The five-dimensional model is the simplest extension of general relativity, and is widely regarded as the low-energy limit of models with even higher dimension (such as 10D supersymmetry and 11D supergravity). Modern versions of 5D or Kaluza-Klein gravity abandon the cylinder and compactification conditions, which caused problems with the cosmological constant and the masses of particles, and consider a large extra dimension. In these theories we experience physics as 4D in nature, because we are either close to spacetime by virtue of it being a special hypersurface (as for membrane theory) or we wander away from it only slowly at a rate governed by the cosmological constant (as for induced-matter theory). In both cases, there is an extra coordinate perpendicular to 4D spacetime. It is unreasonable to expect that all particles have exactly the same value of this (or functions of it). That is, particles in spacetime can in principle have different 5D attributes. These may show up as violations of the WEP. (By analogy, when we look at a modern advertisement panel consisting of light-emitting diodes, we see a 2D surface; but the screen is really 3D, because each diode is connected in the perpendicular direction to a circuit which controls its function.) Conversely we realize that if the WEP is perfectly obeyed then it reflects a <u>symmetry</u> of the 5D geometry. In general, we see that the Weak Equivalence Principle can be regarded as a probe of higher dimensions.

The WEP was appreciated before Einstein used it as a basis for general relativity. It was implicitly tested by Galileo, in his putative experiments where objects of different types were dropped from the tower of Pisa about 1630. It was also verified by Newton, who observed the motion of a simple pendulum in 1686. The torsion balance was developed by Eotvos and used to make an explicit test of the WEP of high accuracy in 1896. The torsion balance was modified by Dicke to include several masses and applied to the problem in 1962. Presently, plans are under way for the Satellite Test of the Equivalence Principle, which should improve the accuracy from order 1 part in $10^{12}$ by a factor of about a million. STEP will use the quiet environment of space to measure the rate of fall in the Earth's field of test objects, chosen to differ in their compositions and other physical properties. The test masses will be designed to mimic gravitational monopoles, in order to reduce the coupling of quadruple and other moments to parts of the satellite. (These would mask a true violation of the WEP, and while differential accelerations due to the lower moments of order 1 part in $10^7$ are expected, they will be averaged out.) The expected accuracy of STEP of 1 part in $10^{18}$ is sufficient to show the "intrusion" of an extra dimension into spacetime, or alternatively confirm the WEP as a near-perfect geometrical symmetry.

To see this, let us consider two results, one to do with the equation of motion for a 5D space and one to do with a "vacuum" solution for it. We use the canonical metric of induced-matter theory [1,2] rather than the warp metric of membrane theory [3,4], though we note that the former theory gives the latter when a brane is inserted and that mathematically the two approaches are essentially the same [5]. We use units which absorb the gravitational constant, the speed of light and Planck's constant and coordinates $x^A = (x^\alpha, l) = (t\ r\ \theta\ \phi, l)$ where $d\Omega^2 \equiv (d\theta^2 + \sin^2\theta\ d\phi^2)$. The 5D line element contains the 4D one via $dS^2 = (l/L)^2 ds^2 - dl^2$, where $L$ is a constant introduced for the consistency of physical dimensions, and $ds^2 = g_{\alpha\beta}(x^\gamma, l) dx^\alpha dx^\beta$ involves a 4D metric tensor which may depend on $x^4 = l$. The 5D line element is mathematically general, in that we have used the 5 degrees of coordinate freedom to remove the electromagnetic potentials ($g_{\alpha 4}$) and flatten the scalar potential ($g_{44}$). This is in accordance physically with our planned applications. The field equations for the 5D metric tensor $g_{AB}$ ($= g_{\alpha\beta}, -1$) are commonly taken in terms of the Ricci tensor to be the apparently empty ones $R_{AB} = 0$ ($A, B = 0\ 1\ 2\ 3, 4$). However, these contain the 4D field equations of general relativity, which in terms of the Einstein tensor and an effective or induced energy-momentum tensor are $G_{\alpha\beta} = 8\pi\ T_{\alpha\beta}$ ($\alpha, \beta = 0\ 1\ 2\ 3$). This embedding is guaranteed by Campbell's theorem [6,7]. The reduction of the 5D field equations to the 4D ones shows that for a 5D metric of the noted form the length $L$ is related to the 4D cosmological constant via $\Lambda = 3/L^2$ (ref. 2, p. 159). With these conventions established, the 5D geodesic equation and the 5D field equations can be solved without much difficulty.

The equations of motion for a minimal path with $\delta [\int dS] = 0$ can be obtained, to give the 5D analogs of "straight" lines. They read

$$\frac{d^2 x^\mu}{ds^2} + \Gamma^\mu_{\alpha\beta} \frac{dx^\alpha}{ds} \frac{dx^\beta}{ds} = f^\mu$$

$$f^\mu \equiv \left(-g^{\mu\alpha} + \frac{1}{2} \frac{dx^\mu}{ds} \frac{dx^\alpha}{ds}\right) \frac{dl}{ds} \frac{dx^\beta}{ds} \frac{\partial g_{\alpha\beta}}{\partial l}$$

$$\frac{d^2 l}{ds^2} - \frac{2}{l}\left(\frac{dl}{ds}\right)^2 + \frac{l}{L^2} = -\frac{1}{2}\left[\frac{l^2}{L^2} - \left(\frac{dl}{ds}\right)^2\right] \frac{dx^\alpha}{ds} \frac{dx^\beta}{ds} \frac{\partial g_{\alpha\beta}}{\partial l} \quad .$$

The first of these relations is the standard equation for geodesic motion (where the Christoffel symbol $\Gamma^{\mu}_{\alpha\beta}$ depends on the usual potentials), but modified by a fifth force per unit mass or acceleration. The latter depends on the motion of the 4D frame with respect to spacetime, and is so inertial in the Einstein sense. The rate of change of the extra coordinate ($x^4 = l$) with respect to the 4D proper time (s) is given by the last of the above relations. If it is zero, then the extra force is zero, as expected. However, $f^{\mu} = 0$ also follows if $\partial g_{\alpha\beta} / \partial l = 0$. That is, the 4D motion is geodesic as in general relativity if the 4D part of the metric does not depend on the extra coordinate [8]. In other words: <u>the Weak Equivalence Principle of 4D can be understood as a geometrical symmetry in 5D</u>.

The field equations can be solved, to give the 5D analogs of the "vacuum" (deSitter) solution. Thus

$$dS^2 = \frac{l^2}{L^2}\{A^2 dt^2 - B^2 dr^2 - C^2 r^2 d\Omega^2\} - dl^2$$

$$A \equiv \left(1 - \frac{r^2}{L^2}\right)^{1/2} + \frac{\alpha L}{l}, \quad B \equiv \frac{1}{(1 - r^2/L^2)^{1/2}}, \quad C \equiv 1 + \frac{\beta L^2}{rl} \quad .$$

Here $\alpha$ and $\beta$ are dimensionless constants, and the length $L$ as we noted above measures the curvature due to the cosmological constant ($\Lambda = 3/L^2$). We therefore recognize a class of solutions, which for $\alpha = 0 = \beta$ contains the deSitter one commonly taken as defining the vacuum in general relativity [9]. In other words: the unique vacuum of 4D becomes multiple vacua in 5D.

We have examined other consequences of the preceding relations and the nature of the extra coordinate elsewhere [1,2,8,9]. But it is apparent from the short discussion given here that whatever the nature of the fifth dimension it can have significant effects on 4D physics. These are presumably small, since we know that general relativity is in excellent agreement with observations [10]. Among planned experiments, the Satellite Test of the Equivalence Principle is of particular interest to us. This will in theory be able to detect departures from the standard laws of motion and effects which might follow from a non-standard vacuum, and will in practice have unprecedented accuracy. While detailed calculations remain to be carried out, STEP has the potential to show us the existence of at least one higher dimension, or confirm that we should be satisfied with spacetime.

Acknowledgements

The work reported here is the result of a group effort, involving contributions from H.Liu, B. Mashhoon, J.M. Overduin and S.S. Seahra.